\documentclass[showpacs,prb,twocolumn]{revtex4}

\usepackage[T1]{fontenc}

\usepackage[thinspace,squaren,pstricks,italian,derivedinbase,derived]{SIunits}
\addunit{\bohr}{au}
\addunit{\rydberg}{Ry}
\addunit{\chwilka}{ch}
\addunit{\kubo}{Kb}

\usepackage[latin2]{inputenc}
\usepackage[dvips]{epsfig}
\usepackage{amsfonts}
\usepackage{amsmath}
\usepackage{amssymb}
\usepackage{upgreek}  
\usepackage{mathrsfs}
\usepackage{booktabs} 
\usepackage{setspace}

\renewcommand{\vec}[1]
{
{\mathbf #1}
}

\newcommand{\Bm}
{
{\mu_{\textrm{B}}}
}

\newcommand{\onemat}
{
{\sf I}     
}

\renewcommand{\Im}
{
{\mathfrak{Im}}
}

\newcommand{\fbr}[1]
{
{\left( #1 \right)} 
}

\newcommand{\xc}
{
_{\mathrm{xc}}
}
\newcommand{\KS}
{
_{\mathrm{KS}}
}

\newcommand{\el}
{%
\textit{et~al}.\
}

\begin{document}


\title{Life times and chirality of spin-waves in antiferromagnetic and ferromagnetic FeRh: time depedent density functional theory perspective} 
\author{Leonid M.\ Sandratskii}
\email[Corresponding author: ]{lsandr@mpi-halle.mpg.de}
\affiliation{Max-Planck-Institut f\"ur Mikrostrukturphysik, Weinberg 2, 06120 Halle (Saale), Germany}
\author{Pawe\l{} Buczek}
\affiliation{Max-Planck-Institut f\"ur Mikrostrukturphysik, Weinberg 2, 06120 Halle (Saale), Germany}
\date{\today}

\begin{abstract}
The study of the spin excitations in antiferromagnetic (AFM) and ferromagnetic (FM) phases of FeRh is reported. We demonstrate that although the Fe atomic moments are well defined there is a number of important phenomena absent in the Heisenberg description: Landau damping of spin waves, large Rh moments induced by the AFM magnons, the formation of the optical magnons terminated by Stoner excitations. We relate the properties of the spin-wave damping to the features of the Stoner continuum and compare the chirality of the spin excitations in AFM, FM and paramagnetic (PM) systems. 
\end{abstract}


\maketitle
 
Antiferromagnets (AFM) form a rich class of magnetically ordered systems characterized by the coexistence of non-zero \textit{local} magnetization and zero \textit{net} magnetic moment. In the contemporary physics, the interests in these materials has been refuelled by experiments on the ultra-fast laser induced magnetization reorientation \cite{Kimel2004}, exchange-bias effect \cite{Morales2009} and the recent discovery of the exotic pnictide family of superconductors. \cite{Mazin2009} The latter materials are often characterized by the proximity of the itinerant AFM and high-temperature superconductivity and the understanding of the spin excitations in the parent AFM compounds is crucial. \cite{Ke2011}

Here, we study the spin-flip excitations in FeRh. There are important reasons for focusing on this system. It is antiferromagnetic at low temperatures and experiences the transition to the ferromagnetic (FM) state at \unit{370}{\kelvin}. This allows the comparison of the AFM and FM spin excitations in the same material. Interestingly, the AFM-FM transition in FeRh can be initiated on the femtosecond time scale by the laser irradiation. \cite{Ju2004}

The microscopic nature of the phase transformation in FeRh remains the topic of controversial debates. It is widely recognized that the crucial role in the stabilization of the FM phase is played by the properties of the Rh atoms whose spin moments increase from zero in the ground AFM state to $\unit{1}{\Bm}$ in the FM phase. Ju \el \cite{Ju2004} suggested that the driving force of the AFM-FM transition is the appearance of the Rh moments in the AFM phase resulting from the fluctuations of the Fe moments. On the other hand, Gu and Antropov \cite{Gu2005} put forward a principally different model where the appearance of the Rh moments in the AFM is not regarded and the phase transition is driven by the difference in the temperature dependence of the spin-wave entropy in both magnetic phases. Recently, we have shown that the spectrum of spin excitations in FeRh is more complex than considered previously. \cite{Sandratskii2011} In particular we demonstrated that strong spin polarizability of the Rh atoms is the consequence of the implicit spin-polarization of the Rh electron states in the AFM ground state.
 
All previous studies of the magnetic excitations in FeRh were performed within adiabatic approaches that map the system on a model Hamiltonian of interacting atomic moments. \cite{Ju2004,Gu2005,Sandratskii2011} A serious disadvantage of the method is the neglect of the one-electron Stoner excitations that can lead to such important effects as damping, and even disappearance, of the spin waves as well as to strong modification of the magnon energies. \cite{Costa2004a} The spin waves and Stoner excitations are incorporated on an equal footing is the calculations of the transverse dynamic susceptibility within the framework of the time dependent density functional theory (TDDFT). \cite{Gross1985,Savrasov1998,Buczek2009} Such studies for AFM systems are very scarce \cite{Edwards1978,Ke2011} because of the demanding character of the underlying computations. Recently, we have developed and implemented an efficient computational scheme for high accuracy calculations of the transverse magnetic susceptibility of complex magnets on the basis of linear response (LR) TDDFT. \cite{Buczek2010d,Buczek2011,Buczek2011a} Here, we report the study of the spin-flip excitations in AFM and FM phases of FeRh.

On the experimental side, there is only one old measurement of the spin wave properties of FeRh performed by means of inelastic neutron scattering for both AFM and FM phases. \cite{Castets1977} The experiment revealed spin waves well defined in the whole Brillouin zone (BZ), in contrast to the \textit{spin-wave disappearance} observed for large wave vectors in elemental $3d$ ferromagnetic bcc Fe \cite{Mook1973,Savrasov1998,Buczek2011a} Additionally, the experimental error bars for the spin-wave energies are very small, and in the case of AFM, the presented error bars do not increase with increasing magnon momentum, contrary to the FM phase where certain increase is seen. In both AFM and FM phases only one spin-wave branch was detected. 
 
The calculation of the dynamic spin susceptibility within LR-TDDFT involves two steps. \cite{Gross1985,Buczek2011,Buczek2011a} Using the ground state Kohn-Sham electronic structure as the input, the non interacting magnetic response, $\chi\KS(\vec{q},\omega)$, of the system is found. The susceptibility does not give the physical magnetic response of the system, since the induced magnetization, leading to the additional induced internal exchange-correlation magnetic field described by the exchange-correlation kernel $K\xc$ \cite{Buczek2011a}, is not taken into account. Since the induced magnetization contributes to the effective external field and is, at the same time, induced by this field, the problem must be solved self-consistently. The physical susceptibility $\chi(\vec{q},\omega)$ is found from the Dyson equation 
\begin{align}
  \chi = \fbr{\onemat - \chi\KS K\xc}^{-1}\chi\KS.
\label{eq:SusceptibilityDysonEquationMatrix}
\end{align}
The imaginary part of the susceptibility gives the intensity of spin-flip excitations with given energy $\omega$ and crystal momentum $\vec{q}$. The information on the single-electron spin-flip spectrum, i.e.\ \textit{Stoner excitations}, is contained in the Kohn-Sham response, $\chi\KS$. The excitations form \textit{Stoner continuum}. The spin-waves form at frequencies corresponding to the vanishing eigenvalues of the matrix $\onemat - \chi\KS K\xc$. \cite{Buczek2011a} If the spin-wave frequency appears outside the Stoner continuum, it features an infinite life-time. On the contrary, the magnons with energies within the continuum decay due to the hybridization with single particle excitations. This attenuation mechanism, \textit{Landau damping}, dominates in metals. Inside the continuum, depending on its intensity, the physical picture of magnons varies between well defined quasi-particles (with life time much longer that their inverse energy) and spin-wave disappearance. We analyze the susceptibility by evaluating the \textit{loss matrix}, $(\chi - \chi^{\dagger})/(2i)$. Its eigenvalues describe the intensity of spin-flip excitations, while the corresponding eigenvectors give the shape of natural modes of the system, i.e.\ spin-waves. \cite{Buczek2011a}

\begin{figure}[htbp]
  \centering
  \includegraphics[width=0.49\textwidth]{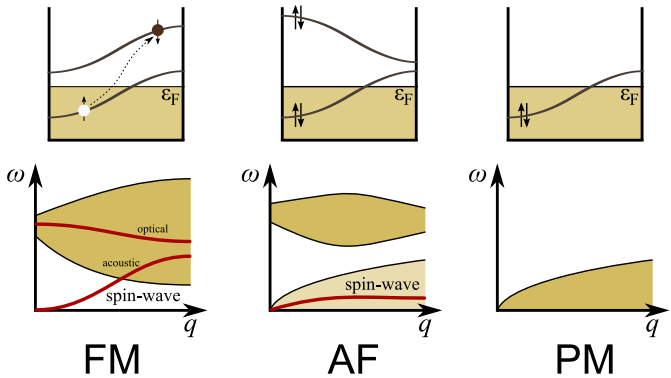}
  \caption{Schematics of band structures and Stoner continua in FM, AFM and paramagnetic material.}
  \label{fig:StonerContinua}
\end{figure}

\begin{figure}[htbp]
  \centering
  a)~\includegraphics[width=0.3\textwidth]{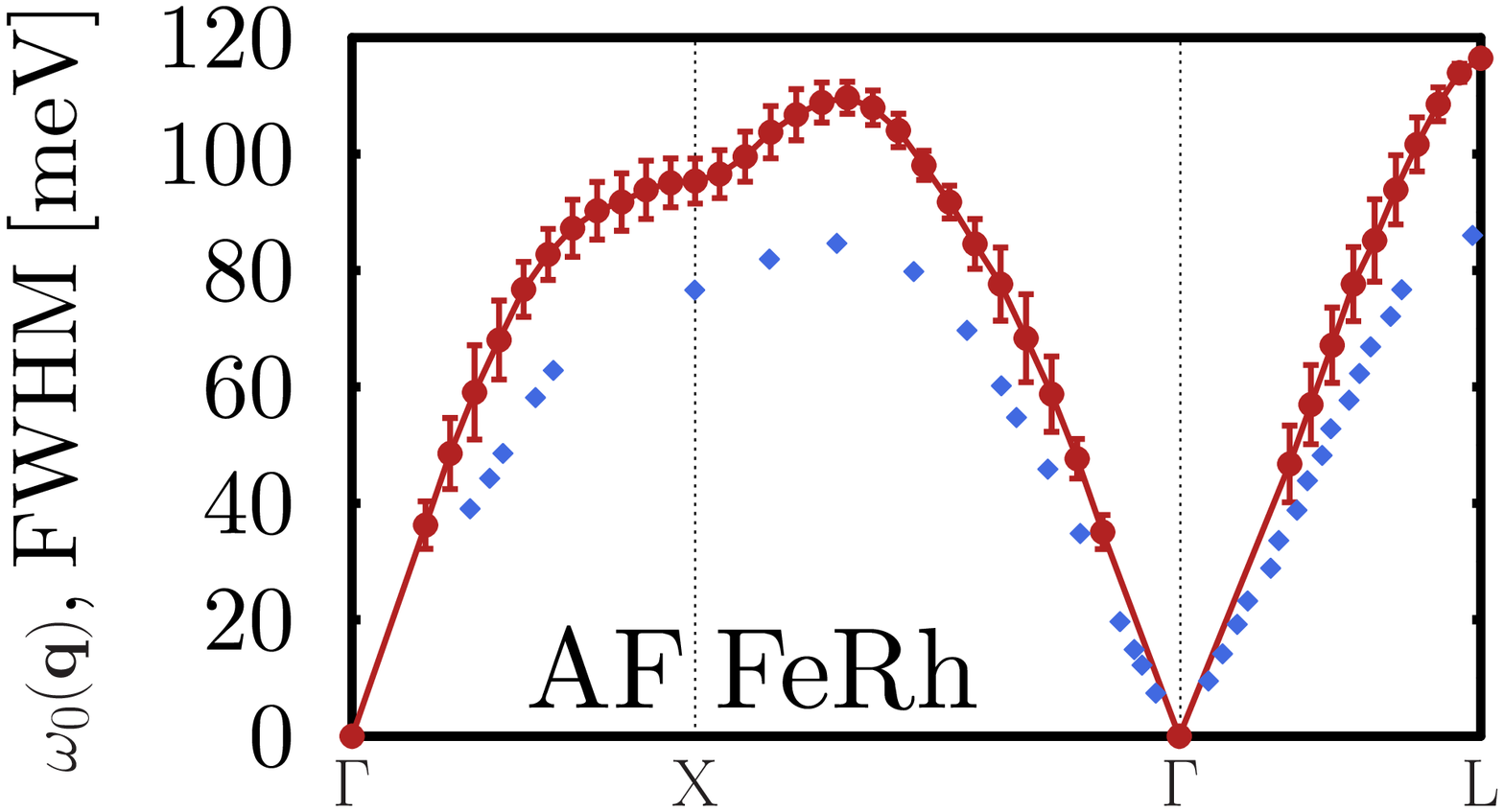}
  b)~\includegraphics[width=0.3\textwidth]{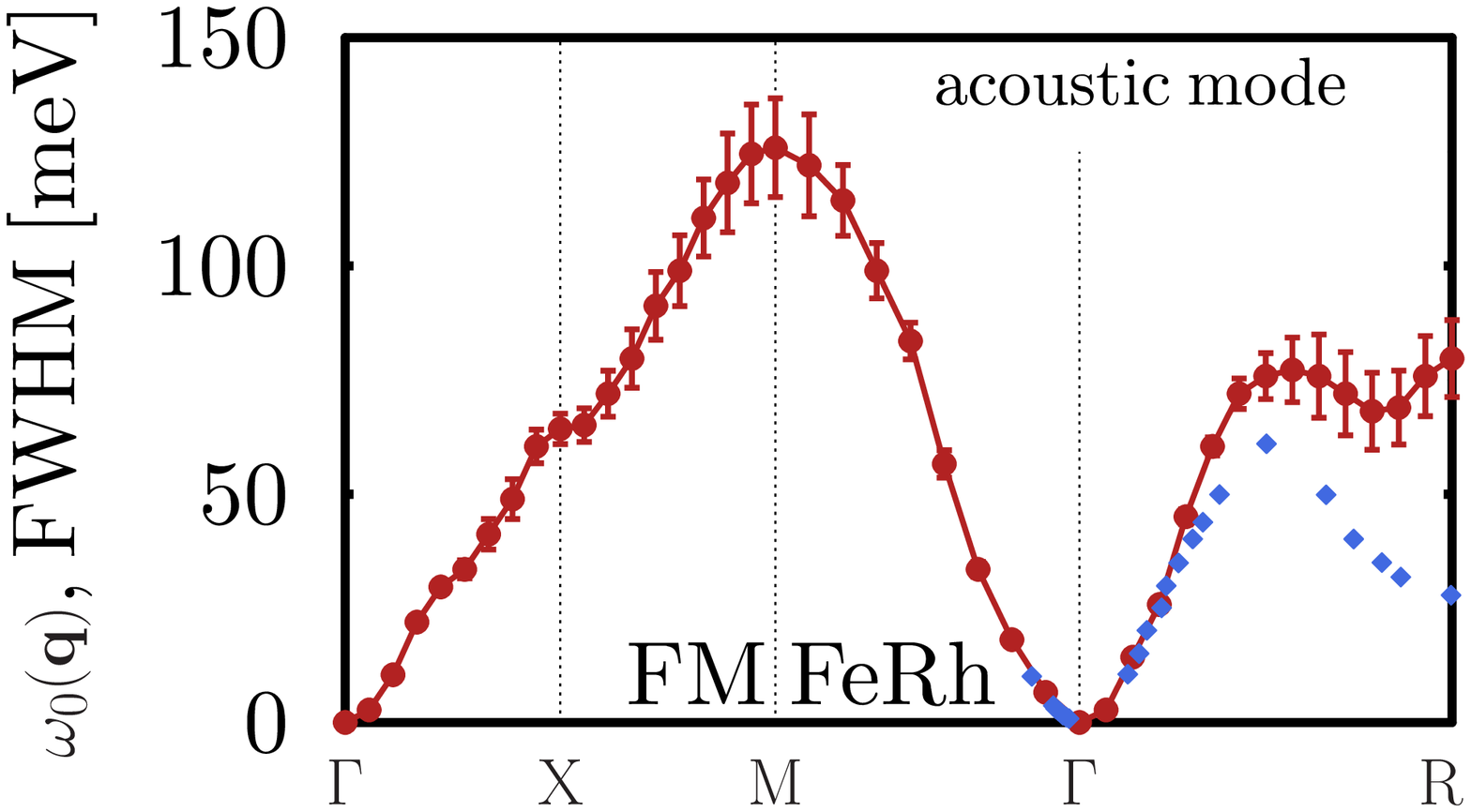}
  \includegraphics[width=0.4\textwidth]{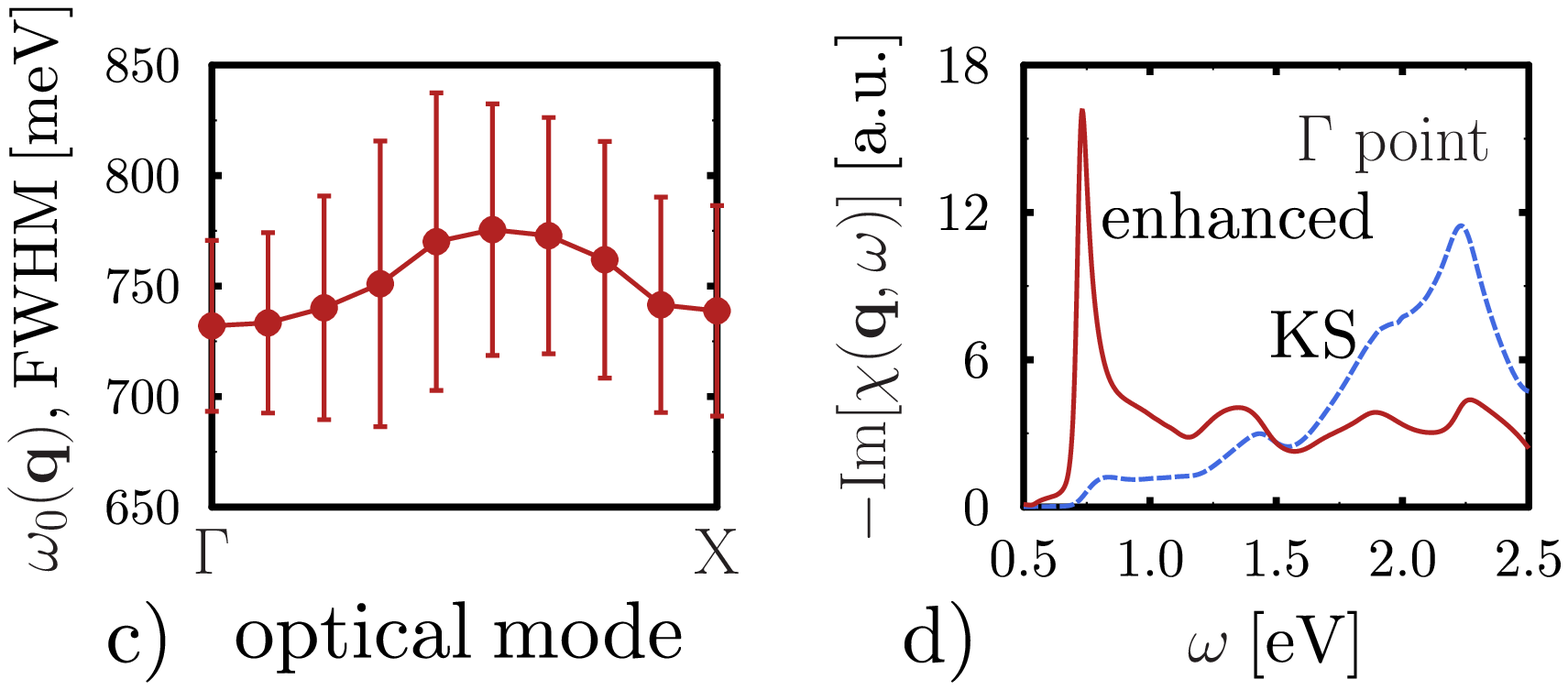}
  \caption{Dispersion relation of spin-waves in the (a) AFM and (b,c) FM phase of FeRh. The dots (\textbullet) represent energy of the magnon peak maximum, while the error bars stand for the full width at half maximum (FWHM) of the peak. Diamonds ($\blacklozenge$) denote the experimental spin-wave energies.\cite{Castets1977} (d) Example of spin-flip spectra in FM FeRh in the energy range of optical magnons.}
  \label{fig:Dispersions}
\end{figure}

This consideration is of a very general nature \cite{Buczek2011a} and is valid for both AFM and FM and even for paramagnets featuring long-living collective spin excitations (paramagnons). There are, however, important qualitative differences in the structure of the Stoner continuum in AFM, FM and paramagnetic (PM) metals. They are illustrated schematically in Fig.\ \ref{fig:StonerContinua} for a simple case of two bands corresponding to two opposite spin projections. The characteristic feature of the FM is the absence of the Stoner transitions in the low-$\omega$, low-q region. This is a consequence of the exchange splitting of the spin-up and spin-down bands. Magnons with small momenta are not Landau damped. On the other hand, in a metallic AFM the Stoner continuum starts at $q=0$ and $\omega=0$ which follows from the spin degeneracy of the electron states. The AFM structure leads to the decrease of the BZ volume compared to the FM and PM cases and the  formation of the second electron band. The transitions between states of the first and second bands form a second area of the high energy Stoner continuum. Also in PM crystal the states are spin-degenerate resulting in the low energy Stoner transitions. There is however an important difference in the character of spin degeneracy in the AFM and PM crystals. In the AFM, the spin degeneracy is the consequence of the presence of two equivalent magnetic sublattices and the wave functions of the degenerated states are shifted in the space with respect to each other. On the other hand, in the PM crystal the wave functions of the degenerate states are identical. Because of the larger overlap of the initial and final states the intensity of the Stoner transitions in the low $q$-$omega$ region tends to be stronger in the PM crystal. For real multiple-band systems the structure of the Stoner continuum is more complex, but the qualitative features depicted in Fig.\ \ref{fig:StonerContinua} remain adequate.

\begin{figure}[htbp]
  \centering
  \includegraphics[width=0.3\textwidth]{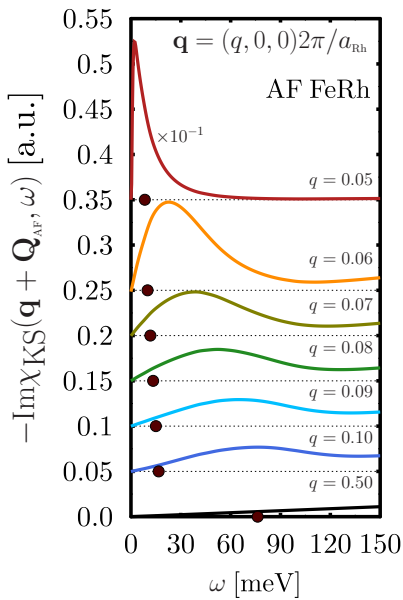}
  \caption{Stoner continuum of AFM FeRh for small momenta. By selecting the $\vec{q} + \vec{Q}_{\mathrm{AF}}$ Fourier component of the KS susceptibility, $\vec{Q}_{\mathrm{AF}} = (1,1,1)2\pi/a_{\mathrm{Rh}}$, we focus on the transitions between spin-degenerate bands. The spin-wave energies are marked with {\Large\textbullet}.}
  \label{fig:SmallQStoner}
\end{figure}

Now, we turn to the discussion of the results of our calculations. The obtained dispersion and damping of the spin waves of AFM FeRh are in good agreement with experiment, cf.\ Fig.\ \ref{fig:Dispersions}a. The damping is not only very small over the whole BZ but it also clearly decreases with increasing momentum above $\unit{80}{\milli\electronvolt}$. In Fig.\ \ref{fig:Dispersions}a we do not show the values of full width at half maximum (FWHM) for the wave vectors close to $\Gamma$, since a reliable numerical estimation in this region is difficult. Instead, we make some qualitative remarks. The velocity of the electronic states at the Fermi level exceeds by an order of magnitude the magnon velocity given by the slope of the magnon dispersion at the $\Gamma$ point. This means that already for small momenta the magnons form inside the Stoner continuum and therefore for any $q\ne0$ their damping is non-zero. The analysis of the Kohn-Sham susceptibility (Fig.\ \ref{fig:SmallQStoner}) shows that for small $q$ the spectral density of the Stoner continuum has a shape of a narrow peak with its width vanishing in the $q\to0$ limit. This is a consequence of the decrease of energy differences between electron states involved in the transitions at small momenta. On the other hand, for increasing $q$ the width of the peak increases and the position of the peak moves to larger frequencies. Since the energy width of the peak of the enhanced susceptibility determining the position and life time of the spin wave cannot be larger than the energy area of the non-zero Stoner continuum the damping tends to zero in the $q\to0$ limit. Therefore, the spin-wave branch begins at $q=0$ with infinitely-long living magnons of vanishing energy. For $q>0$ the spin-wave peaks acquire non-zero width. On the basis of Fig.\ \ref{fig:SmallQStoner} we can conclude that the damping in the $q$-interval from 0 to 0.1 first increases and than decrease again because of strongly decreasing spectral weight of the Stoner transitions at the spin wave energy.

\begin{figure}[htbp]
  \centering
  \includegraphics[width=0.3\textwidth]{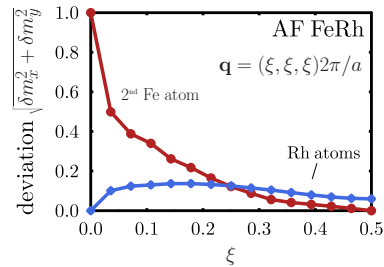}
  \caption{Transverse magnetization induced on the second Fe lattice and Rh site, assuming the deviation of the first Fe lattice is normalized to one.}
  \label{fig:Deviations}
\end{figure}

\begin{figure}[htbp]
  \centering
  \includegraphics[width=0.3\textwidth]{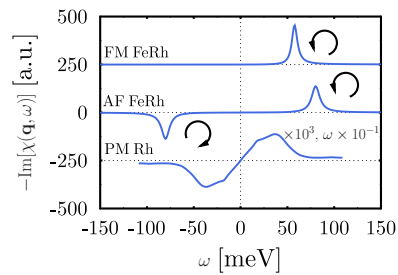}
  \caption{Chirality of spin-waves in different magnetic phases. Arrows denote the direction of magnetization precession.}
  \label{fig:Chirality}
\end{figure}

As mentioned in the introduction, the spin moments of the Rh atoms play very important role in the magnetism of FeRh. The calculation of the dynamic spin susceptibility allows to investigate the spin-polarization of the Rh atoms by the AFM magnons. This question is in the direct relevance to the model of femtosecond generation of magnetization suggested Ju \el. \cite{Ju2004} The analysis of the loss matrix shows that indeed such a polarization takes place. To understand this effect it is essential to recall that an AFM magnon breaks the equivalence of the two magnetic sublattices. \cite{Keffer1953} In order for the spins on both sublattices to precess coherently, the moments of one of the sublattices must deviate from the ground state direction stronger than the moments of the other. The ratio of the deviation angles of the two sublattices is $q$ dependent. This is demonstrated in Fig.\ \ref{fig:Deviations} where we show the transverse magnetization induced on the second Fe lattice and Rh site, assuming the deviation of the first Fe lattice is normalized to one. For $q\to0$ the ratio tends to 1 what is a necessary condition for the fulfillment of the Goldstone theorem. On the other hand, at the BZ boundary the spins of the atoms of only one sublattice deviate from the ground state direction. The equivalence of the sublattices is restored by the presence of two degenerate spin wave branches corresponding to similar excitations with respect to different sublattices.

In the AFM spin-wave the compensation of the Fe exchange fields at the Rh sites is disturbed. This leads to the appearance of non-zero atomic moments at the Rh sites. The value of the moments induced by the spin-wave is also $q$ dependent, cf.\ Fig.\ \ref{fig:Deviations}. It is zero for $q\to0$ but assumes sizable values of 18\% of the transverse Fe moments for larger momenta. Although Fe atomic moments in FeRh are well defined, the richness of the physics of FeRh transcends the Heisenberg model incapable to describe the magnetization induced on Rh sites.

In the FM (Fig. \ref{fig:Dispersions}) we have very good agreement with experiment in the low $q$ region whereas closer to the zone boundary the deviation between theory and experiment becomes bigger. The damping has a clear tendency to increase with increasing momentum. A sizable damping is obtained for only rather large magnon wave vectors. This correlates with the property of the electronic structure of the FM FeRh that has very small number of the Fe spin-up states in the nearest vicinity of the Fermi level. \cite{Sandratskii2011}

Since in FM the Rh atoms have a large moment of $\unit{1}{\Bm}$, it is natural within the Heisenberg model to consider them as Heisenberg variables that leads to two spin wave modes: acoustic and optical. \cite{Gu2005} At higher frequencies we found sharp features in the enhanced spectral density, cf.\ Fig.\ \ref{fig:Dispersions}d. The analysis of the eigenvectors of the loss matrix shows that the polar angle of the Rh magnetization deviation increases by $180^\circ$ with respect to the direction of the Rh moment in the acoustic magnon. This is a characteristic feature of an optical magnon. The spectral feature, however, is characterized by a strong damping and appears at the energy position of abruptly increasing Stoner continuum, below the expected energy of the optical magnon. The resulting peak can be interpreted neither as spin-wave nor as single-particle Stoner excitations, having a complex mixed character. The position and width of the resonance as a function of wave vector is presented in Fig.\ \ref{fig:Dispersions}. In principle, these resonances should be observable in the inelastic neutron scattering experiment for sufficiently large neutron energies and momenta. \cite{Buczek2010a} The peaks, however, carry much smaller spectral weight than the corresponding  acoustic resonances.  

For comparison, we performed calculations for paramagnetic Rh. As expected the Stoner continuum starts in this case from the zero energy (not shown), similar to the case of AFM. It is however much more intense, in particular, because of the the stronger overlap of the wave functions of the states involved in the transitions. The enhanced susceptibility repeats the main features of the KS one. The enhancement in this material is not sufficiently strong to form a band of well defined paramagnon excitations similar to that found , e.g., in Pd. \cite{Staunton2000} Since in paramagnetic metals the spin density vanishes, the symmetry with respect to the spin rotation is not spontaneously broken and the Goldstone mode does not form. In contrast to the spin waves in the AFM, the collective excitations in the PM are of longitudinal character and are connected with the formation of the non-zero magnetization instead of its rotation.

The comparison of the spectral densities of the spin-flip excitations for a given $q$ value in the cases of FM, AFM and PM reveals interesting similarities and differences between them. In the FM, there is an asymmetry with respect to the change of the sing of the frequency: the spin-wave can be excited only at positive frequencies, which is a consequence of the broken time reversal symmetry of the FM ground state. In contrast, in the AFM the spectral density is symmetric with respect to the change of the sign of the frequency. These are two magnon branches that can be transformed to each other by the symmetry operation consisting of the product of the time reversal and space translation transforming one sublattice into other. By this transformation the sublattices interchange their roles. The paramagnet is invariant with respect to the time reversal that leads to the chirality property of the transverse magnetization similar to the case of AFM.
   
To summarize, we show that although the Fe atomic moments are well defined in FeRh there are a number of important phenomena that are missed by the Heisenberg model: Landau damping of spin waves, large Rh moments induced by the AFM magnons, the termination of the formation of the optical magnons by Stoner excitations. AFM, FM and PM systems differ strongly with respect to the chirality of their collective spin-flip excitations. We hope this theoretical work will stimulate experiments on the spin-flip excitations in FeRh, e.g.\ using modern high-flux neutron sources.


\end{document}